\documentclass[showpacs,preprintnumbers,amsmath,amssymb]{elsart}
\usepackage{rotating}
\usepackage{graphicx}
\usepackage{color}
\usepackage{bm}
\usepackage{epsfig}
\begin{document}
\bibliographystyle{try}
\newcommand{\er}{$\pm$}
\newcommand{\be}{\begin{eqnarray}}
\newcommand{\ee}{\end{eqnarray}}
\newcommand{\widt}{\rm\Gamma_{tot}}
\newcommand{\wadd}{$\rm\Gamma_{miss}$}
\newcommand{\gpiN}{$\rm\Gamma_{\pi N}$}
\newcommand{\getN}{$\rm\Gamma_{\eta N}$}
\newcommand{\gkla}{$\rm\Gamma_{K \Lambda}$}
\newcommand{\gksi}{$\rm\Gamma_{K \Sigma}$}
\newcommand{\gNpi}{$\rm\Gamma_{P_{11} \pi}$}
\newcommand{\gDpf}{$\rm\Gamma_{\Delta\pi(L\!<\!J)}$}
\newcommand{\gDps}{$\rm\Gamma_{\Delta\pi(L\!>\!J)}$}
\newcommand{\sqgDpf}{$\rm\sqrt\Gamma_{\Delta\pi(L\!<\!J)}$}
\newcommand{\sqgDps}{$\rm\sqrt\Gamma_{\Delta\pi(L\!>\!J)}$}
\newcommand{\gnsi}{$\rm N\sigma$}
\newcommand{\gNpf}{$\rm\Gamma_{D_{13}\pi}$}
\newcommand{\gNps}{$\Gamma_{D_{13}\pi(L\!>\!J)}$}
\newcommand{\roper}{$\rm N(1440)P_{11}$}
\newcommand{\srma}{$\rm N(1535)S_{11}$}
\newcommand{\trma}{$\rm N(1520)D_{13}$}
\newcommand{\srmb}{$\rm N(1650)S_{11}$}
\newcommand{\trmb}{$\rm N(1700)D_{13}$}
\newcommand{\trmc}{$\rm N(1875)D_{13}$}
\newcommand{\trmd}{$\rm N(2170)D_{13}$}
\newcommand{\fvma}{$\rm N(1675)D_{15}$}
\newcommand{\fvmb}{$\rm N(2070)D_{15}$}
\newcommand{\fvpa}{$\rm N(1680)F_{15}$}
\newcommand{\srpb}{$\rm N(1710)P_{11}$}
\newcommand{\trpa}{$\rm N(1720)P_{13}$}
\newcommand{\trpb}{$\rm N(2200)P_{13}$}
\newcommand{\trpd}{$\rm N(2170)D_{13}$}
\newcommand{\dtpa}{$\rm\Delta(1232)P_{33}$}
\newcommand{\doma}{$\rm\Delta(1620)S_{31}$}
\newcommand{\dtma}{$\rm\Delta(1700)D_{33}$}
\newcommand{\dtmb}{$\rm\Delta(1940)D_{33}$}
\newcommand{\rthe}{A^{1/2}/A^{3/2}}
\newcommand{\amoh}{$A^{1/2}$}
\newcommand{\amth}{$A^{3/2}$}
\newcommand{\broh}{\Gamma^{1/2}_{\gamma p}/\widt}
\newcommand{\brth}{\Gamma^{3/2}_{\gamma p}/\widt}
\newcommand{\btot}{$\Gamma(\gamma p)$}
\newcommand{\Dpi}{\Delta\pi}
\newcommand{\KL}{\rm\Lambda K}
\newcommand{\KS}{\rm\Sigma K}

\newcounter{univ_counter}
\setcounter{univ_counter} {0}
\addtocounter{univ_counter} {1}
\edef\HISKP{$^{\arabic{univ_counter}}$ } \addtocounter{univ_counter}{1}
\edef\GATCHINA{$^{\arabic{univ_counter}}$ } \addtocounter{univ_counter}{1}
\edef\BASEL{$^{\arabic{univ_counter}}$ } \addtocounter{univ_counter}{1}
\edef\GIESSEN{$^{\arabic{univ_counter}}$ } \addtocounter{univ_counter}{1}
\edef\MAINZ{$^{\arabic{univ_counter}}$ } \addtocounter{univ_counter}{1}
\edef\GLASGOW{$^{\arabic{univ_counter}}$ }
\addtocounter{univ_counter}{1}
\edef\ERLANGEN{$^{\arabic{univ_counter}}$ } \addtocounter{univ_counter}{1}
\edef\FSU{$^{\arabic{univ_counter}}$ } \addtocounter{univ_counter}{1}
\edef\PI{$^{\arabic{univ_counter}}$ } \addtocounter{univ_counter}{1}
\edef\BOCHUM{$^{\arabic{univ_counter}}$ } \addtocounter{univ_counter}{1}
\edef\KVI{$^{\arabic{univ_counter}}$ } \addtocounter{univ_counter}{1}

\begin{frontmatter}

\title{New results on the Roper resonance and the
\boldmath$P_{11}$\unboldmath\ partial wave}

\author[HISKP,GATCHINA]{A.V.~Sarantsev},
\author[HISKP]{M.~Fuchs},
\author[BASEL,GIESSEN]{M.~Kotulla},
\author[HISKP,GIESSEN]{U.~Thoma},
\author[MAINZ]{J.~Ahrens},
\author[GLASGOW]{J.R.M.~Annand},
\author[HISKP,GATCHINA]{A.V.~Anisovich},
\author[ERLANGEN]{G.~Anton},
\author[PI]{R.~Bantes},
\author[HISKP]{O.~Bartholomy},
\author[HISKP,MAINZ]{R.~Beck},
\author[GATCHINA]{Yu.~Beloglazov},
\author[KVI]{R.~Castelijns},
\author[HISKP,FSU]{V.~Crede},
\author[HISKP]{A.~Ehmanns},
\author[HISKP]{J.~Ernst},
\author[HISKP]{I. Fabry},
\author[BOCHUM]{H.~Flemming},
\author[ERLANGEN]{A.~F\"osel},
\author[HISKP]{Chr.~Funke},
\author[PI]{R.~Gothe},
\author[GATCHINA]{A.~Gridnev},
\author[HISKP]{E.~Gutz},
\author[PI]{St.~H\"offgen},
\author[HISKP]{I.~Horn},
\author[ERLANGEN]{J.~H\"o\ss l},
\author[MAINZ]{D.\,Hornidge},
\author[GIESSEN]{S.\,Janssen},
\author[HISKP]{J.~Junkersfeld},
\author[HISKP]{H.~Kalinowsky},
\author[PI]{F.~Klein},
\author[HISKP]{E.~Klempt},
\author[BOCHUM]{H.~Koch},
\author[PI]{M.~Konrad},
\author[BOCHUM]{B.~Kopf},
\author[BASEL]{B.~Krusche},
\author[PI]{J.~Langheinrich},
\author[KVI]{H.~L\"ohner},
\author[GATCHINA]{I.~Lopatin},
\author[HISKP]{J.~Lotz},
\author[GLASGOW]{J.C.~McGeorge},
\author[GLASGOW]{I.J.D.~MacGregor},
\author[BOCHUM]{H.~Matth\"ay},
\author[PI]{D.~Menze},
\author[GIESSEN,KVI]{J.G.\,Messchendorp},
\author[GIESSEN]{V.~Metag},
\author[HISKP,GATCHINA]{V.A.~Nikonov},
\author[GATCHINA]{D.~Novinski},
\author[GIESSEN]{R.~Novotny},
\author[PI]{M.~Ostrick},
\author[HISKP]{H.~van~Pee},
\author[GIESSEN]{M.~Pfeiffer},
\author[GATCHINA]{A.~Radkov},
\author[GLASGOW]{G.\,Rosner},
\author[MAINZ]{M.\,Rost},
\author[HISKP]{C.~Schmidt},
\author[PI]{B.~Schoch},
\author[ERLANGEN]{G.~Suft},
\author[GATCHINA]{V.~Sumachev},
\author[HISKP]{T.~Szczepanek},
\author[PI]{D.~Walther},
\author[GLASGOW]{D.P.\,Watts}, and
\author[HISKP]{Chr.~Weinheimer}\\

\address[HISKP]{Helmholtz-Institut f\"ur Strahlen- und Kernphysik der
Universit\"at Bonn, Germany}
\address[GATCHINA]{Petersburg
Nuclear Physics Institute, Gatchina, Russia}
\address[BASEL]{Physikalisches Institut, Universit\"at
Basel, Switzerland}
\address[GIESSEN]{II. Physikalisches Institut, Universit\"at Giessen}
\address[MAINZ]{Institut f\"ur Kernphysik, Universit\"at Mainz, Germany}
\address[GLASGOW]{Department of Physics and Astronomy, University of
Glasgow, UK}
\address[ERLANGEN]{Physikalisches Institut, Universit\"at
Erlangen, Germany}
\address[PI]{Physikalisches Institut, Universit\"at Bonn, Germany}
\address[KVI]{KVI, Groningen, Netherlands}
\address[FSU]{Department of Physics, Florida State
University, USA}
\address[BOCHUM]{Physikalisches Institut, Universit\"at
Bochum, Germany}

\date{\today}

 \clearpage


\begin{abstract}
Properties of the Roper resonance, the first scalar excitation of the
nucleon, are determined. Pole positions and residues of the $P_{11}$
partial wave are studied in a combined analysis of pion- and
photo-induced reactions. We find the Roper pole at
$\{(1371\pm7)-i(92\pm10)\}$\,MeV and an elasticity of $0.61\pm 0.03$.
The largest decay coupling is found for the $N\sigma$
($\sigma=(\pi\pi)$-$S$-wave).
%
The analysis is based on new data on $\gamma p\to p\pi^0\pi^0$ for
photons in the energy range from the two-pion threshold to 820\,MeV
from TAPS at Mainz and from 0.4 to 1.3\,GeV from Crystal Barrel at
Bonn and includes further data from other experiments. The partial
wave analysis excludes the possibility that the Roper resonance is
split into two states with different partial decay
widths. \vspace{2mm}   
\end{abstract}

\end{frontmatter}
\vspace{-3mm}

The lowest-mass excitation of the nucleon, the Roper $N(1440)P_{11}$
resonance with spin and parity $J^P=1/2^+$, and the second scalar
nucleon excitation $N(1710) P_{11}$ \cite{Yao:2006px}, remain to be
the most enigmatic states in baryon spectroscopy. In the bag model
\cite{Meissner:1984un} and in the Skyrme model \cite{Hajduk:1984ry},
the Roper resonance was interpreted as surface oscillation, also
called breathing mode. In quark models, two low-mass scalar
excitations of the nucleon are predicted. Using a linear confining
potential and one-gluon-exchange \cite{Capstick:bm} or
instanton-induced interactions \cite{Loring:2001kx}, a level
ordering is calculated in which the mass of the $N(1440)P_{11}$
exceeds the mass of the negative-parity state $N(1535)S_{11}$ by
80\,MeV; experiments find it $\sim 100$\,MeV below. The spacing
between the two scalar excitations is predicted to be $\sim220$\,MeV
\cite{Capstick:bm,Loring:2001kx,Loring:2001ky} while experiments
find 270\,MeV \cite{Yao:2006px}. When one-gluon exchange
interactions are replaced by exchanges of Goldstone bosons, the
$N(1440)P_{11}$ mass can well be reproduced \cite{Glozman:1997ag},
the $N(1710) P_{11}$ mass was not calculated.  Lattice gauge
calculations indicate that the first scalar excitation of the
nucleon should be expected above $N(1535)S_{11}$
\cite{Burch:2006cc}. Compared to model and lattice predictions, the
mass of the Roper resonance is too small; compared to other low-mass
resonances, its width too large.

These problems would not occur if $N(1710)P_{11}$ were the first
radial scalar excitation of the proton. The Roper resonance can then
be interpreted within a coupled-channel meson exchange model based
on an effective chiral-symmetric Lagrangian \cite{Krehl:1999km}; no
genuine $N(1440)$ (3 quark) resonance was needed to fit $\pi N$
phase shifts and inelasticities, in agreement with
\cite{Schneider:2006bd}. Motivated by the $Q^2$ dependence of the
Roper helicity amplitude which would seem to suggest a hybrid nature
\cite{Li:1991yb}, Capstick and Page \cite{Capstick:1999qq}
calculated masses of baryonic hybrids. Their masses were, however,
too large to interpret the Roper resonance as a hybrid. The
$\Theta^+(1530)$, a baryon with positive strangeness, which may have
been observed in low-statistics photo-production experiments
\cite{Nakano:2003qx,Barmin:2003vv,Stepanyan:2003qr,Barth:2003es},
made the Roper resonance \cite{Jaffe:2003sg} and/or the
$N(1710)P_{11}$ \cite{Diakonov:1997mm} to viable pentaquark
candidates. The existence of a very narrow $P_{11}$ state in the
mass region 1650-1750 MeV was investigated in~\cite{Arndt:2003ga}.
The fading evidence for $\Theta^+(1530)$
\cite{Battaglieri:2005er,Niccolai:2006td} makes this interpretation
less attractive. Morsch and Zupranski \cite{Morsch:2000xi} suggested
the Roper mass region might house two resonances, one at 1390\,MeV
with a small elastic width and large coupling to $N\pi\pi$, and a
second one at higher mass -- around 1460\,MeV -- with a large
elastic width and small $N\pi\pi$ coupling. The former resonance was
found to be produced in $\pi N$ scattering, and in $\alpha$-proton
scattering using an $\alpha$ beam of $E_{\alpha}=4.2$\,GeV kinetic
energy; the latter resonance was suggested to be excited by $\gamma
N$. The two resonances may have rather different wave functions
\cite{Morsch:2004kn}. Studies of the reaction $pp\to pp\pi^+\pi^-$
suggested that the low-energy tail of the Roper resonance might
decay to both $N\sigma$ and $\Delta\pi$ \cite{Patzold:2003tr}.
Obviously, the $P_{11}$ partial wave is not sufficiently constrained
by precision data and supports very different interpretations.

In this letter, we present data on $\gamma p \to p\pi^0\pi^0$ of the
A2-TAPS collaboration at the Mainz Microtron (MAMI) electron
accelerator \cite{ahrens:mami} and of the CB-ELSA collaboration at
the Bonn ELectron Stretcher Accelerator (ELSA)
\cite{Hillert:2006yb}. The Bonn set up and the analysis method was
described briefly in the preceding paper \cite{Thoma:2006}. Here we
give only a short summary of the TAPS  setup, for more details see
\cite{kottu:2pi0thres}. Earlier data taken at MAMI
\cite{Harter:1997jq,Wolf:2000qt} have smaller statistics and are not
discussed here.

\begin{figure}[pt]
\begin{minipage}[c]{0.42\textwidth}
\hspace{-2mm}\epsfig{file=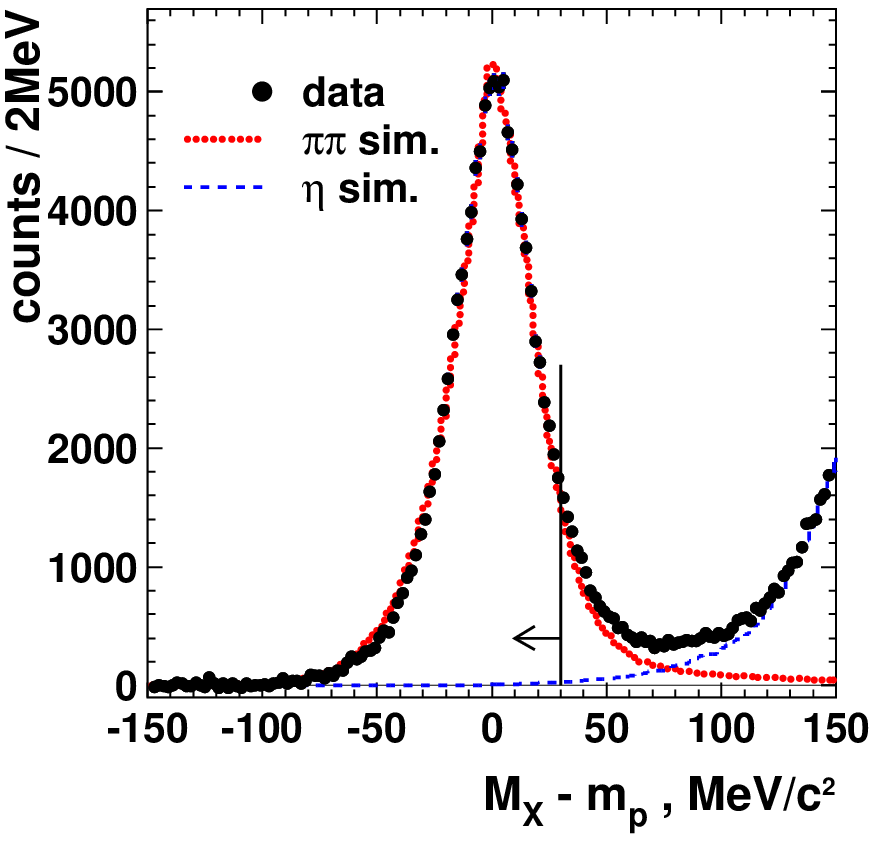,width=1.0\textwidth,height=6.2cm,clip=}
\end{minipage}
\begin{minipage}[c]{0.56\textwidth}
\hspace{-3mm}\epsfig{file=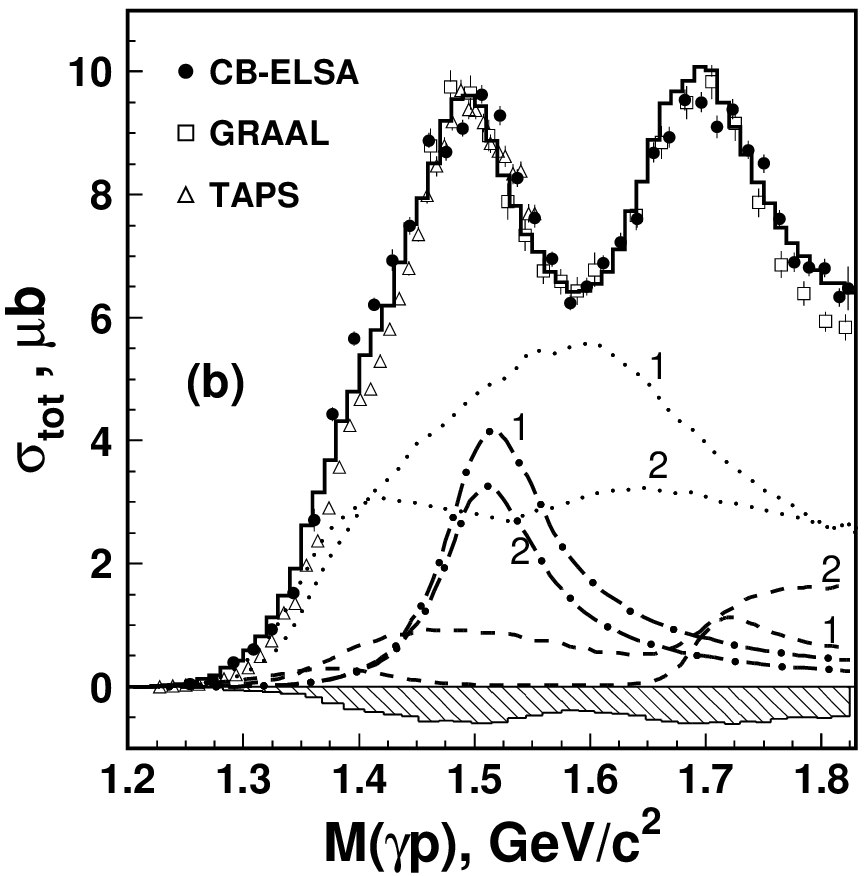,width=1.06\textwidth,height=6.0cm,clip=}
\end{minipage}
\caption{\label{gg_vs_gg} a) TAPS: Missing mass $M_X - m_p$ calculated
from two detected $\pi^0$ mesons for incident beam energies
$E_\gamma$ $\leq$ 820 MeV (data: symbols with errors, $\pi^0\pi^0$
simulation dotted line, $\eta$ simulation dashed line). The cut to
eliminate $\eta$ background is indicated. b) Total cross sections
for $\gamma p \to p \pi^0\pi^0$. The shaded area below the zero line
represents the systematic error of the CB-ELSA data, the solid line a
PWA fit. There are two PWA solutions, marked 1 and 2, giving a similar
likelihood (see text). The $D_{33}$ partial wave (dotted line) gives
the strongest contribution to the second resonance region, followed by
$D_{13}$ (dashed-dotted line) and $P_{11}$ (dashed line). The $D_{13}$
-- $D_{33}$ interference generates the dip between the second and third
resonance region.
 }
\end{figure}

The photon energy at MAMI covered the range 285--820\,MeV. The photon
energies were measured in the Glasgow tagged photon facility
\cite{hall:tagger} with an average energy resolution of 2\,MeV. The
TAPS detector \cite{novotny:taps,gabler:response} consisted of six
blocks each with 62 hexagonally shaped BaF$_2$ crystals arranged in an
8$\times$8 matrix and a forward wall with 138 BaF$_2$ crystals arranged
in a 11$\times$14 rectangle. This setup covered $\approx$40\% of
$4\pi$. The $\gamma p \rightarrow \pi^0 \pi^0 p$ reaction channel was
identified by constructing the 4-momenta of the two neutral pions from
their $\gamma\gamma$ decays; proton detection was not required in the
analysis. The $\pi^0$ mesons were detected via their $2\gamma$ decay
and identified by their invariant mass. The mass of the missing
particle was calculated from the four-momenta of the pions, and the beam
energy $E_{\gamma}$ using the mass of the target proton. The resulting
distribution is shown in Fig.~\ref{gg_vs_gg}a and demonstrates the
unambiguous identification of the reaction $\gamma p \rightarrow \pi^0
\pi^0 p$. At incident beam energies above the $\eta$ production
threshold of $E_\gamma = 707$~MeV, a possible background from the $\eta
\rightarrow 3\pi^0$ decay with one undetected $\pi^0$ can be cleanly
separated from the reaction of interest (see Fig.~\ref{gg_vs_gg}a).
Further details are given in \cite{kottu:mdm_prl}.

In Fig.~\ref{gg_vs_gg}b the total  cross section is displayed. Two peaks due
to the second and third resonance region are observed, with peaks at
$\sim 1500$ and $\sim 1700$\,MeV. There is good general agreement
between the three data sets. The GRAAL data \cite{Assafiri:mv} fall off
at high masses more rapidly than the CB-ELSA data. At low energies, the
TAPS data fall below the CB-ELSA data while the peak cross sections of
all 3 experiments agree reasonably well. The discrepancies show the
difficulties of extracting total cross sections when the full phase
space is not covered by the detector. Note that the extrapolation was
done differently: the CB-ELSA and the A2-TAPS collaborations used the
result of this partial wave analysis; the GRAAL collaboration used a
simulation based on $\gamma p\to \Delta^+\pi^0$ and $\gamma p\to
p\pi^0\pi^0$ phase space. The inclusion of both, CB-ELSA and TAPS data,
provides an additional tool to estimate the systematic error of the
experimental data. The fit curves in Fig.~\ref{gg_vs_gg}b are discussed
below.

\begin{figure}[pt]
\begin{tabular}{ccc}
\epsfig{file=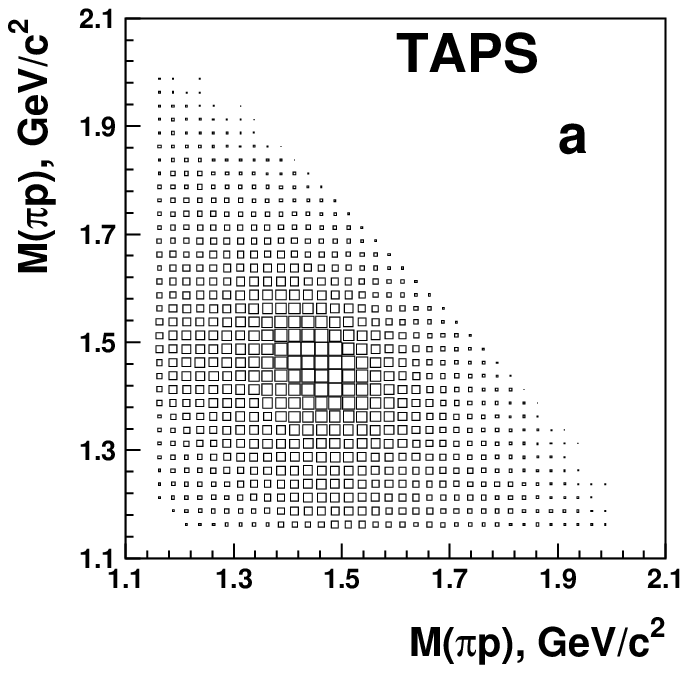,width=0.32\textwidth,clip=}&
\epsfig{file=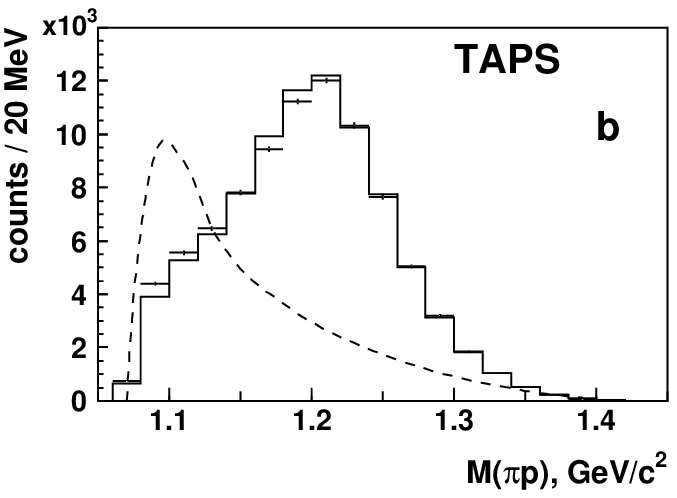,width=0.32\textwidth,height=0.32\textwidth,clip=}&
\epsfig{file=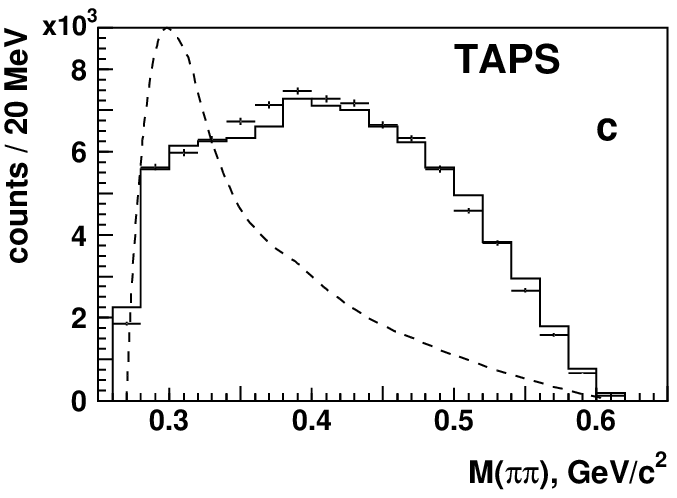,width=0.32\textwidth,height=0.32\textwidth,clip=}\\
\epsfig{file=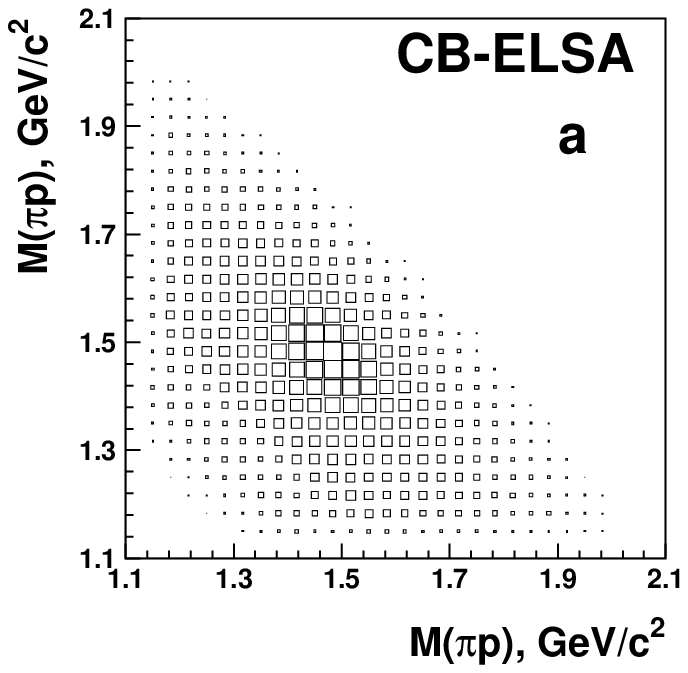,width=0.32\textwidth,clip=}&
\epsfig{file=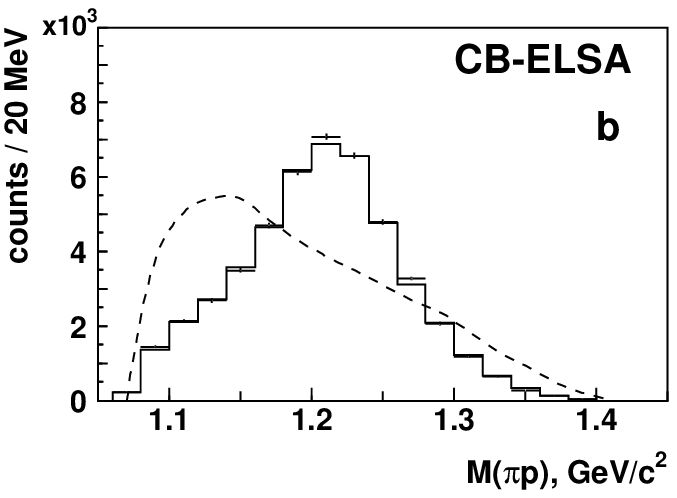,width=0.32\textwidth,height=0.32\textwidth,clip=}&
\epsfig{file=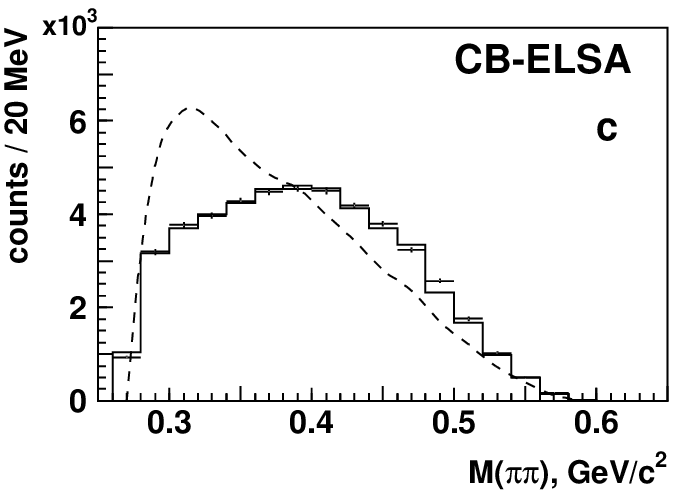,width=0.32\textwidth,height=0.32\textwidth,clip=}\\
\end{tabular}
\caption{\label{masang}
Dalitz plots (a) and $p\pi^0$ (b) and $\pi^0\pi^0$ (c)
mass distributions for TAPS (upper row) and for CB-ELSA (lower row) data
for $\gamma p\to p\pi^0\pi^0$. The CB-ELSA data shown here are
restricted to $M\leq 1.55$\,GeV. In (b,c), data are represented by
crosses, the fit as histogram. The dashed lines represents the phase
space contributions within the acceptance. The distributions are not
corrected for acceptance which are different for the two experiments,
leading to different distributions. }
\end{figure}

The total cross section gives only a very superficial view of the
reaction. Fig.~\ref{masang}a shows the experimental $p\pi^0\pi^0$
Dalitz plots, Fig.~\ref{masang}b,c the $p\pi^0$ and $\pi^0\pi^0$ mass
distributions. The solid line represents the result of a fit, the
dashed line represents the distribution of reconstructed phase space
events. The projections are not corrected for detection efficiency to
allow the reader to compare data and fit directly. From the $\pi^0 p$
mass distributions we conclude that the $\Delta$ isobar plays an
important role in the two-pion photoproduction dynamics. The
$\pi^0\pi^0$ mass distributions are featureless but show strong deviations
from phase space.

Even the Dalitz plot and the projections do not carry the full
information on the reaction dynamics. The full sensitivity of the data
can only be exploited using an event-based likelihood fit. The data
presented here were subjected to a partial wave analysis based on the
Bonn-Gatchina approach \cite{Anisovich:2004zz,Anisovich:2006bc}.
Compared to a previous analysis
\cite{Anisovich:2005tf,Sarantsev:2005tg}, several new data sets are
included in this analysis. A list of additional data and a description
of the partial wave analysis method can be found in \cite{cxcz}. In the
case of two-particle final states (including $\gamma p\to\Lambda K^+$
and $\gamma p\to\Sigma K$), angular distributions are fitted;
three-body final states like $N\pi^0\pi^0$ and $p\pi^0\eta$
\cite{Horn:2007} undergo an event-based likelihood fit.

\begin{table}[pb]
\begin{center}
\caption{\label{reactions}The reactions most important for the
study of properties of the Roper resonance.\vspace{2mm}}
\begin{tabular}{cccc} \hline\hline 1 & $\gamma
p\to p\pi^0\pi^0$&Figs. \ref{gg_vs_gg},\ref{masang}&this work\\ 2 &
$\gamma p\to p\pi^0$ & Figs. \ref{cb-pi0}-\ref{r-pol} &
\cite{Bartholomy:2004uz,van Pee:2007tw,Bartalini:2005wx,Dugger:2007bt} \\ 3 & $\pi N
\to N\pi$ & Fig. \ref{p11-amp}& \cite{Arndt:2006bf}              \\ 4 &
$\pi^- p\to n\pi^0\pi^0$ & Fig. \ref{ball} & \cite{Prakhov:2004zv}\\
\hline\hline
\end{tabular} \end{center}
\end{table}

\begin{figure}[ph]
\begin{center} \epsfig{file=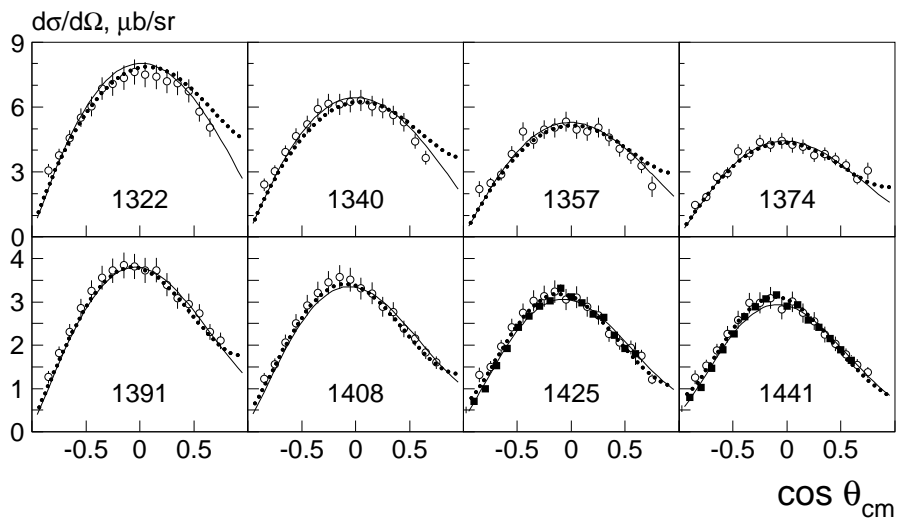,width=0.86\textwidth,clip=}
\end{center}
\caption{\label{cb-pi0} The $\gamma p\to p\pi^0$ differential cross
section. Open circles \cite{Bartholomy:2004uz}; full
circles:\cite{Bartalini:2005wx} . Solid lines: our fit; dotted
lines: SAID, the solution FA06 \cite{Dugger:2007bt}.}
\end{figure}

\begin{figure}[ph]
\begin{center}
\epsfig{file=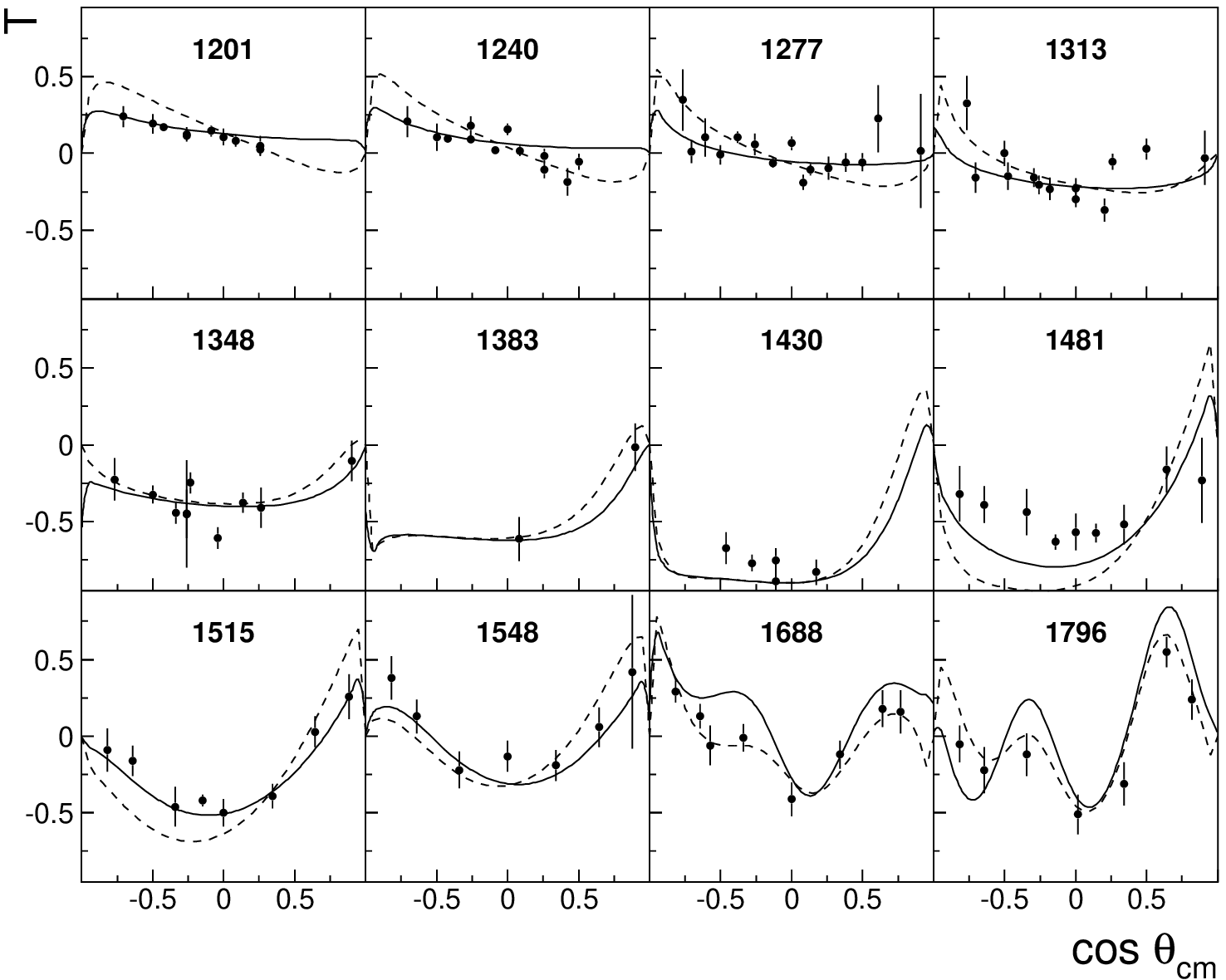,width=0.85\textwidth,clip=}
\end{center}
\caption{\label{t-pol} The target asymmetry from different
experiments \cite{SAID} for selected  5\,MeV mass bins. Solid lines:
our fit; dashed lines: SAID, the solution FA06 \cite{Dugger:2007bt}.
}\vspace{3mm}
\begin{center}
\epsfig{file=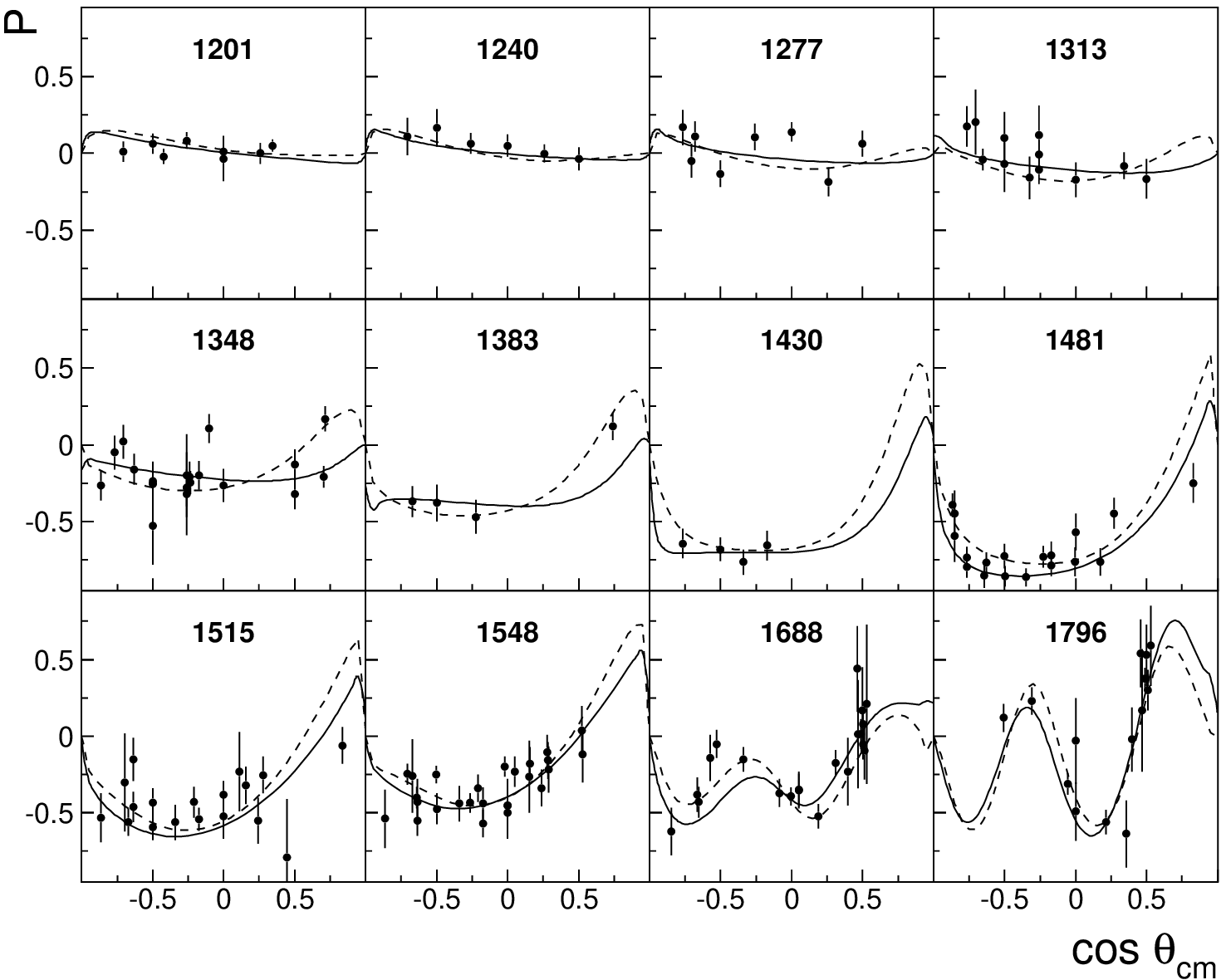,width=0.85\textwidth,clip=}
\end{center}
\caption{\label{r-pol} The proton recoil polarization from different
experiments \cite{SAID} for selected  5\,MeV mass bins. Solid lines:
our fit; dashed lines: SAID, the solution FA06 \cite{Dugger:2007bt}.
}
\end{figure}

\begin{figure}[ph]
\begin{center}
\epsfig{file=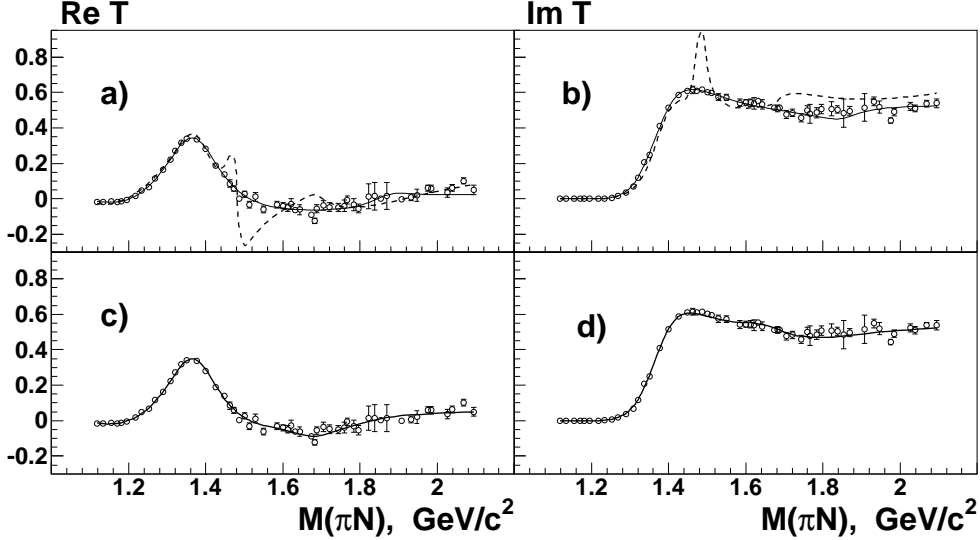,width=0.95\textwidth,clip=}
\end{center}
\caption{\label{p11-amp}Real (a,c) and imaginary (b,d) part of the
$\pi$N $P_{11}$ elastic scattering amplitude; data and fit with two
(a,b) and three (c,d) poles \cite{Arndt:2006bf}. The dashed line in
(a,b) represents a fit in which the Roper resonance is split into
two components. The overall likelihood deteriorates to extremely bad
values. The fit tries to make one Roper resonance as narrow as
possible. }
\end{figure}

\begin{figure}[pt]
\begin{tabular}{ccc}
\hspace{-2mm}\epsfig{file=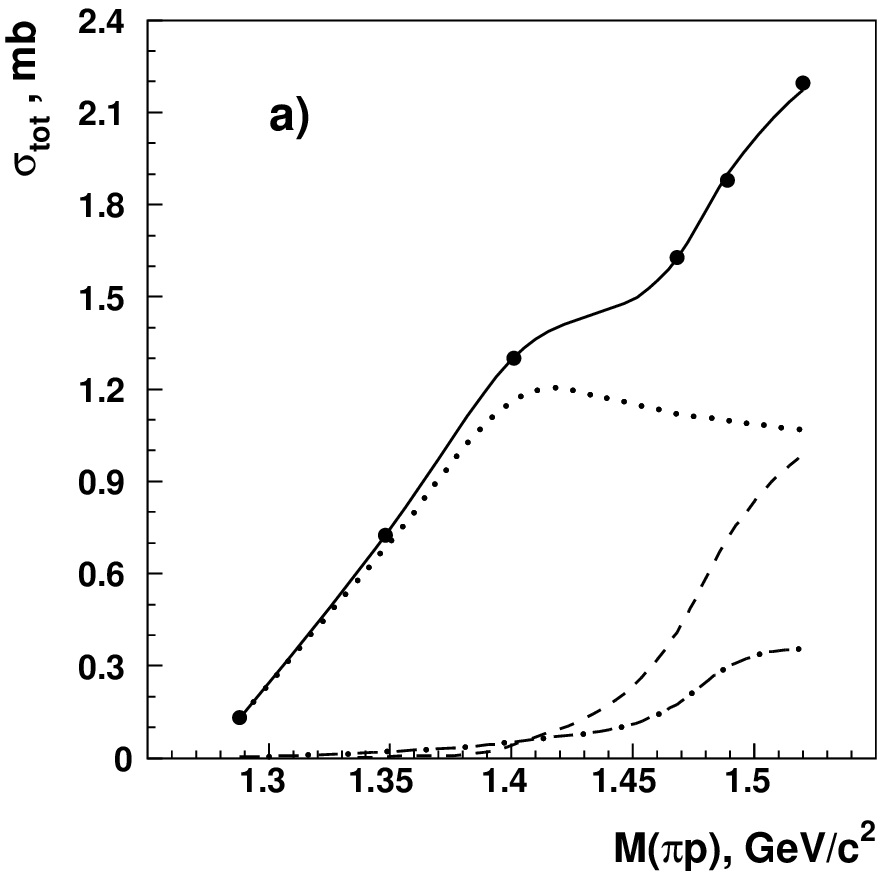,width=0.32\textwidth,height=0.2\textheight,clip=}&
\hspace{-2mm}\epsfig{file=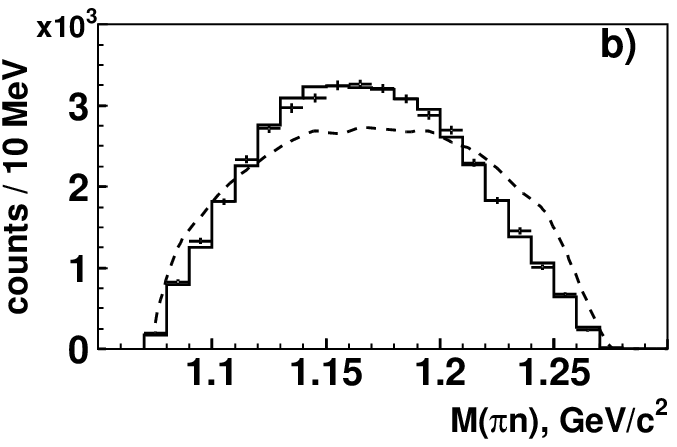,width=0.32\textwidth,height=0.2\textheight,clip=}&
\hspace{-2mm}\epsfig{file=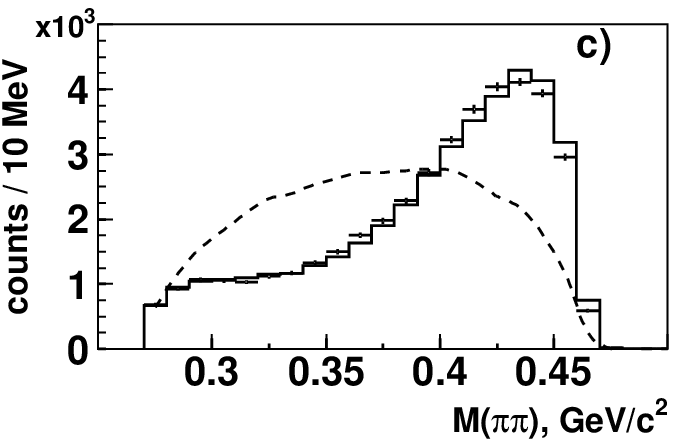,width=0.32\textwidth,height=0.2\textheight,clip=}
\end{tabular}
\caption{\label{ball} The reaction $\pi^-p\to n\pi^0\pi^0$
\cite{Prakhov:2004zv}. (a) Total cross section; the errors are
smaller than the dots. (b) $\pi^0 P$ and (c) $\pi^0\pi^0$ invariant
mass distributions for 551\,MeV/c. In (a) the dotted, dashed and
dashed dotted lines give the $P_{11}$, $D_{13}$, and $S_{11}$
contributions, respectively. In (b,c) data (crosses), fit
(histogram) and phase space (dashed line) are shown. The
distributions are not corrected for acceptance. }
\end{figure}

The reactions most relevant for the present analysis are collected
in Table~\ref{reactions}. The $\pi N$ elastic scattering
\cite{Arndt:2006bf} amplitude provides a strong constraint for
$N\pi$ partial decay widths of resonances in this partial wave. The
inclusion of data on $\gamma p \to p\pi^0$
\cite{Bartholomy:2004uz,van
Pee:2007tw,Bartalini:2005wx,Dugger:2007bt} and on $\pi^-p\to
n\pi^0\pi^0$ \cite{Prakhov:2004zv} over-constrains resonance
properties: the three partial decay widths, $\Gamma_{N\gamma}$,
$\Gamma_{N\pi}$, and $\Gamma_{N\pi\pi}$, of the Roper
$N(1440)P_{11}$ resonance have to describe its properties in four
reactions. In addition, the partial widths define the number of
events ascribed to the Roper resonance in the reactions above. Their
sum must equal the total width since the only missing channel,
$N(1440)P_{11}\to N\rho$, is expected to provide a very small
contribution due to the small available phase space. We believe that
the tight constraints due to the use of four different reactions
defines the Roper mass, width, and coupling constants with much
higher reliability than analyses of individual reactions can do. The
background amplitudes in the four reactions are treated
independently.  The data and the quality of the description are
shown in Figs.~\ref{gg_vs_gg}-\ref{ball}.

We started the partial wave analysis from the solution given in
\cite{Anisovich:2005tf,Sarantsev:2005tg}. Including the new data, we
found good compatibility for masses and widths of the contributing
resonances. The new description of single $\pi^0$ photoproduction is
shown in Fig. \ref{cb-pi0}. In Figs. \ref{t-pol} and \ref{r-pol} we
present a comparison of fit and data on target asymmetry and on the
proton recoil polarization from different experiments \cite{SAID}.
Inclusion of the latter data had an impact on the size of couplings
but did not change the pole positions; the properties of the Roper
resonance were nearly unaffected. The figures (not the fits) are
restricted to the mass range below 1800\,MeV.

The $\gamma p\to p\pi^0\pi^0$ data provide new information on the
N$\pi\pi$ decay modes. The $D_{33}$ amplitude gives the largest
contribution to this data (see Fig.~\ref{gg_vs_gg}b). Its
interference with $N(1520)D_{13}$, constructive at $\sim$1500\,MeV,
destructive at $\sim$1600\,MeV, generates the dip between the two
peaks in the total cross section. \\The $\Delta(1700)D_{33}$ has a
large coupling to $\Delta\pi$. In the main solution,
$\Delta(1700)D_{33}$ decays into $\Delta\pi$ in a relative $S$ wave.
There is however a second solution with very similar likelihood in
which the $\Delta(1700)D_{33}\to\Delta\pi$ decays proceed via

\clearpage
$D$-wave. This ambiguity results in different contributions of all
$p\pi^0\pi^0$ partial waves as shown in Fig.~\ref{gg_vs_gg}b. The
interference between $\Delta(1700)D_{33}$ and background
contributions is responsible for the shallow dent in solution 2 of
the $D_{33}$ contribution visible in Fig.~\ref{gg_vs_gg}b.

The importance of the $(\pi\pi)$-$S$-wave was already hinted at by
Murphy and Laget (quoted by the GRAAL collaboration in
\cite{Assafiri:mv}), although in a very different framework based on
an effective isobaric Lagrangian. In the Laget model, the Roper
resonance provided the largest contribution to $N\pi\pi$, followed
by the $D_{13}$ partial wave while $D_{33}(1700)$ was very weak. The
Valencia model \cite{GomezTejedor:1995pe,Nacher:2000eq} is limited
to $E_{\gamma}<0.84$\,GeV; it predicted strong contributions of
$D_{13}(1520)$ and small $N(1440)P_{11}$ and $D_{33}(1700)$
contributions. Our analysis finds a very strong $D_{33}(1700)$
contribution. However, the dominant orbital angular momentum in the
$D_{33}(1700)\to \Delta\pi$ decay is ambiguous giving rise to two
acceptable solutions. Both solutions are fully compatible with the
$D_{33}$ $\pi N$ elastic scattering amplitude (see Fig.~5 in
\cite{Horn:2007}). The analysis presented here is constrained by a
large number of data sets and exploits all two-particle correlations
within the $N\pi\pi$ final state. These technical differences may
very well be a reason for the discrepant results.

The $P_{11}$ amplitude for $\pi N$ elastic scattering is written in the
form of a K-matrix containing three constants, describing non-resonant
contributions to elastic and inelastic reactions, and a series of poles
representing resonant contributions. The $P_{11}$ photoproduction
amplitude is written as a K-matrix in P-vector approach (which neglects
$p\gamma$ loops in the rescattering series). The photoproduction
amplitude has the same poles as the scattering matrix. One constant
each is introduced for reaction (1) and (2) in Table
\ref{reactions} describing direct $p\pi^0$ and $p\pi^0\pi^0$
production. The cons\-tants, pole posi\-tions
and couplings $g_{Nx}$ to a final state $N\,x$ are free para\-meters of
the fit. The Born term is described by a pole at the proton mass. At
least two poles were required, with pole positions at 1370 and
1850\,MeV, respectively. In Fig. \ref{p11-amp}a,b we show the $P_{11}$
amplitude for the two-pole solution. The data are well described.

As a next step, we introduced a second pole in the Roper region, a
pion-induced resonance R and a second photo-induced R'. This attempt
failed. The fit reduced the elastic width to the minimal allowed
value of 50\,MeV; the overall probability of the fit became
unacceptable. The resulting elastic amplitude is shown in Fig.
\ref{p11-amp}a,b as dashed line. We did not find any meaningful
solution where the Roper region could comprise two resonances.

In \cite{Anisovich:2005tf,Sarantsev:2005tg}, no evidence for
$N(1710)P_{11}$ was found. The increased
sensitivity due to new data encouraged us to introduce a third pole
in the $P_{11}$ amplitude. Fig. \ref{p11-amp}c,d show the result of
this fit. A small improvement due to $N(1710)P_{11}$ is observed,
and also other data sets are slightly better described. The
parameters of the resonance are not well defined, the pole position
is found in the 1580 to 1700\,MeV mass range.

Introduction of the $N(1710)P_{11}$ as third pole changes the
$N(1840)P_{11}$ properties. In the two-pole solution, the
$N(1840)P_{11}$ resonance is narrow ($\sim 150$\,MeV), in the
three-pole solution, the $N(1710)P_{11}$ and a $\sim 250$\,MeV wide
$N(1840)P_{11}$ resonance interfere to reproduce the structure. Data
with polarized photons and protons will hopefully clarify existence
and properties of these additional resonances. Further $P_{11}$
poles are expected at larger masses. Introducing such a pole does
not lead to a significant improvement of the fit.

The properties of the $N(1440)P_{11}$ resonance determined here are
listed in Table \ref{1440}. From the K-matrix poles and their
couplings, the poles of the scattering matrix T were deduced. The
speed plot $|dT|/dm$ gives $M_{\rm speed}\sim 1340$\,MeV. The
Breit-Wigner parameters are deduced by the following method. The
helicity coupling and the coupling constant for a given decay mode
are calculated as residues of the T-matrix pole in the corresponding
complex $s=M^2$ plane. These are complex numbers. The partial decay
widths are calculated from the coupling constants and the available
phase space including centrifugal barrier factor and Blatt-Weisskopf
corrections \cite{Anisovich:2004zz}. These partial widths (including
the missing width) are scaled by a common factor to reproduce the
T-matrix pole position. The errors cover the range of a large
variety of different PWA solutions.

\begin{table}[pt]
\caption{\label{1440}
Properties of $N(1440)P_{11}$.
The left column lists mass, width, partial widths of the Breit-Wigner
resonance; the right column pole position and squared couplings to the
final state at the pole position.}\vspace{2mm}
\renewcommand{\arraystretch}{1.4}
\begin{center}
\begin{tabular}{lcccccc}
\hline\hline
M&=&$1436\pm15$\,MeV&\qquad&M$_{\rm pole}$&=&$1371\pm7$\,MeV\\
$\Gamma$&=&$335\pm40$\,MeV&& $\Gamma_{\rm pole}$&=&$192\pm20$\,MeV\\
$\Gamma_{\pi N}$&=&$205\pm25$\,MeV&& $g_{\pi
N}$&=&$(0.51\pm0.05)\cdot e^{-i\pi\frac{(35\pm5)}{180}}$\\
$\Gamma_{\sigma N}$&=&$71\pm17$\,MeV&&$g_{\sigma
N}$&=&$(0.82\pm0.16)\cdot e^{-i\pi\frac{(20\pm13)}{180}}$\\
$\Gamma_{\pi\Delta}$&=&$59\pm15$\,MeV&&$g_{\pi\Delta}$&=
&$(-0.57\pm0.08)\cdot e^{i\pi\frac{(25\pm20)}{180}}$\\
\multicolumn{7}{c}{T-matrix:  $A_{1/2}=
0.055 \pm 0.020$\,GeV \qquad $\phi = (70 \pm 30)^{\circ}$}\\
\hline\hline \end{tabular} \renewcommand{\arraystretch}{1.0}
\end{center}\end{table}

The fractional yields of resonant and non-resonant parts are of
course ill-defined quantities. To allow the reader to appreciate
better the meaning of the results, we have set to zero the resonant
or non-resonant part of the amplitude and calculated the
corresponding cross sections, integrated over the Roper region
(1300-1500\,MeV). Interferences are neglected. The results on the
different photoproduction reactions are presented in Table
\ref{1440_contr}. Resonant and non-resonant contributions are
comparatively large and interfere destructively to yield the
observed $P_{11}$ wave.

\begin{table}[ph]
\caption{\label{1440_contr}
Contributions of the $P_{11}$-wave to different photoproduction
reactions, integrated over the 1300-1500\,MeV mass range.}
\vspace{2mm} \renewcommand{\arraystretch}{1.4}
 \begin{center}
 \begin{tabular}{lccc}
 \hline\hline
Reaction & $P_{11}, obs$ (\%)
         & $P_{11}, res$ (\%)
         & $P_{11}, nonres$ (\%)\\
 \hline
 $\gamma p\to\pi^0 p$            &$2.4\pm 0.8$&$4 \pm 1$     &$ 7\pm 2$  \\
 $\gamma p\to\pi^0\pi^0 p$     &$7\pm 2$    &$6 \pm 2$     &$11\pm 3$  \\
 $\quad\gamma p\to\Delta^+\pi^0$   &$5\pm 1$  &$4 \pm 1$     &$7 \pm 2$  \\
 $\quad\gamma p\to p\sigma$    &$4\pm 1$    &$3\pm 1$      &$7 \pm 2$  \\
 \hline\hline
 \end{tabular}
 \renewcommand{\arraystretch}{1.0}
 \end{center}
 \end{table}

This is different in pion scattering. The largest contribution to
the $n\pi^0\pi^0$ final state goes via the $N(1440)P_{11}$
resonance. The complicated interference between resonant and
non-resonant amplitudes may be the reason why the Roper resonance is
so difficult to identify in photoproduction reactions.

\begin{table}[pt]
\caption{\label{1440_pion}
Fractional contributions (in \%) of the $N(1440)P_{11}$ and its
isobars to $\pi^- p\to p\pi^0\pi^0$ for 3 different $\pi^-$ energies.}
\vspace{2mm}\renewcommand{\arraystretch}{1.4}
 \begin{center}
 \begin{tabular}{lccc}
 \hline\hline
 \qquad $p_{\pi}$ (MeV/c)             & 472 & 551 & 655 \\
 \hline
 $\pi^- p\to P_{11}\to n\pi^0\pi^0$ &$95\pm 3$&$88\pm 3$ & $60\pm 5$\\
 $\pi^- p\to P_{11}\to \Delta^0\pi^0$     &$22\pm 3$   &$29\pm 3$  & $25\pm 3$  \\
 $\pi^- p\to P_{11}\to p\sigma$       &$70\pm 5$   &$53\pm 2$  & $32\pm 6$  \\
 \hline\hline
 \end{tabular}
 \renewcommand{\arraystretch}{1.0}
 \end{center}
 \end{table}

The properties of the Roper derived here are mostly
consistent with previous determinations. Pole position and Breit-Wigner
mass and width fall into the range of values given by the Particle Data
Group (PDG \cite{Yao:2006px}),
\begin{center} \begin{tabular}{rccrc}
\hspace{-2mm}$M_{BW} = 1430-1470$&MeV;&&$\Gamma_{BW}
=250-450$&MeV \\ \hspace{-2mm}$M_{\rm pole}=
1345-1385$&MeV;&&$\Gamma_{\rm pole}=160-260$&MeV
 \end{tabular}
 \end{center}
\noindent but are defined more precisely here. The helicity coupling
agrees with the PDG mean value but from the variety of different
solutions we estimate a larger error. Note that our helicity
amplitude is defined in the complex $s=M^2$ plane at the pole
position of the Roper. The elastic width
($\Gamma_{N\pi}/\Gamma_{tot}= 0.612\pm 0.020$) is compatible with
previous findings (60-70\%). Its decay fraction into $\Delta\pi$
($\Gamma_{\Delta\pi}/\Gamma_{tot}= 0.176\pm 0.020$) is not in
conflict with the PDG mean value (20-30\%); only the $N\sigma$
partial decay width deviates significantly from PDG. We find
$\Gamma_{N\sigma}/\Gamma_{tot}= 0.212\pm 0.030$ while PDG gives
5-10\%.

Due to its larger phase space, decays into $N\pi$ are more
frequent than those into $N\sigma$ even though the latter decay
mode provides the largest coupling. For a radial excitation, this is
not unexpected: about 50\% of all $\psi(2S)$ resonances decay into
J/$\psi\,\sigma$, more than 25\% of $\Upsilon(2S)$ resonances decay
via $\Upsilon(1S)\,\sigma$ \cite{Yao:2006px}. The
large value of $g_{\sigma N}$ might therefore support the interpretation of
the Roper resonance as radial excitation.

An alternative interpretation of the $N\pi\pi$ decay is offered by
Hernandez, Oset and Vicente-Vacas \cite{Hernandez:2002xk} who take
into account the re-scattering of the two final state pions. The authors
of \cite{Hernandez:2002xk} do not fit data; instead they show that
they can reproduce qualitatively the phenomenology of
$N(1440)P_{11}\to N\pi\pi$ decays by rescattering thus avoiding the
need to introduce a genuine $N(1440)P_{11}\to N\sigma$ amplitude.

In this letter, we have presented new data on photoproduction of two
neutral pions in the energy range from the 2$\pi^0$ production
threshold up to a photon energy of 820\,MeV (Mainz) and up to
1300\,MeV (Bonn) and reported results from a partial wave analysis
of this and of related reactions.

The focus of this letter is the Roper resonance. We show that the
data are incompatible with the conjecture that conflicting results
on its properties could originate from the presence of two similar
resonances and their interference, where both are in the $P_{11}$
wave and both fall into the 1300 to 1500\,MeV mass region. Due to
the fact that the Roper properties are over-constrained by the data,
we can rule out this possibility. The decay pattern is consistent
with an interpretation of the Roper resonance as first radial
excitation of the nucleon.

\section*{Acknowledgements} \vspace{-3mm}

We would like to thank the technical staff of the ELSA and MAMI
machine groups and of all the participating institutions of their
invaluable contributions to the success of the experiment. We
acknowledge financial support from the Deutsche
Forschungsgemeinschaft (DFG) within the SFB/TR16 and from the
Schweizerische Nationalfond. The collaboration with St. Petersburg
received funds from DFG and RFBR. U.Thoma thanks for an Emmy Noether
grant from the DFG. A.V.~Sarantsev acknowledges support from RSSF.

\end{document}